# Magnetoresistance and surface roughness study of electrodeposited Ni$_{50}$Co$_{50}$/Cu multilayers

B.G. Tóth*, L. Péter, J. Dégi and I. Bakonyi
*Wigner Research Centre for Physics, Hungarian Academy of Sciences.
H-1525 Budapest, P.O.B. 49, Hungary*
(Apr. 05, 2013)

**Abstract** ─ Room-temperature transport properties (the zero-field resistivity, $\rho_0$, and the GMR) were studied for ED Ni$_{50}$Co$_{50}$/Cu multilayers as a function of the individual layer thicknesses and total multilayer thickness. The Cu deposition potential was optimized in order to obtain the preset layer thicknesses. The surface roughness development was studied by AFM, which revealed an exponential roughening with total thickness. The Cu layer thickness strongly influenced the roughness evolution. As expected, $\rho_0$ decreased with increasing Cu layer thickness whereas it increased strongly for large total multilayer thicknesses that could be ascribed to the observed deposit roughening. All multilayers with Cu layer thicknesses above about 1.5 nm exhibited a GMR behavior with a maximum GMR of about 5 %. The GMR decreased for total multilayer thicknesses above about 300 nm due to the strong increase of $\rho_0$, the latter caused by the enhanced roughness. The GMR data indicated the appearance of a current-at-angle-to-plane type scattering due to the layer undulations. The thickness evolution of the MR data was analyzed in detail after separating the ferromagnetic and superparamagnetic GMR contributions. It could be established that ED Ni-Co/Cu multilayers do not exhibit an oscillatory GMR behavior with spacer thickness.

---

*Corresponding author. E-mail: toth.bence@wigner.mta.hu

## Introduction

The giant magnetoresistance (GMR) effect in nanoscale ferromagnetic/non-magnetic (FM/NM) metallic multilayers has widespread applications today.[1-13] Whereas the GMR effect was originally discovered in Fe/Cr multilayers,[14,15] practical application could only be realized by using TM/Cu type multilayers where TM stands for a binary alloy among the elements of the iron-group metals (Fe, Co and Ni) or sometimes Co alone. The reason for this is that only some specific TM/Cu multilayers fulfill simultaneously the requirement for a sufficiently large GMR effect and low saturation field ($H_s$) around room temperature. Namely, the final figure of merit of a GMR sensor is determined by the magnetic field sensitivity measured by the ratio $GMR/H_s$. The GMR effect is the largest for Co/Cu multilayers at the first antiferromagnetic (AF) maximum (about 60 % for a Cu spacer thickness of 0.9 nm) which is associated, however, with a saturation field of about 4 to 5 kOe due to the strong AF exchange coupling between adjacent magnetic layers.[16-18] On the other hand, although the *GMR* magnitude is reduced by a factor of two at the second AF maximum (for a Cu spacer thickness of about 2 nm), due to the much weaker exchange coupling here,[16-18] the sensitivity is already sufficiently large for practical applications. In some magnetically soft binary and ternary alloys containing Fe, Co and/or Ni, one can also find a compromise between *GMR* magnitude and saturation field provided a sufficiently weak coupling can be achieved by properly choosing the magnetic layer composition and spacer layer thickness. A detailed GMR study of sputtered TM/Cu multilayers by using binary and ternary alloys of Fe, Co and Ni as magnetic layers has been carried out by Miyazaki and coworkers[19,20] who also mapped out the anisotropic magnetoresistance (*AMR*) of bulk alloys of the Fe-Co-Ni system.[20] The composition dependence of *GMR* was investigated in particular for sputtered Ni-Co/Cu multilayers by Kubota et al.[18] and Bian et al.[21,22]

In addition to the physical deposition methods used for the preparation of the above described GMR multilayers, electrodeposition was also demonstrated to be capable of yielding multilayers in the TM/Cu systems with sufficiently large GMR effect.[23,24] As to the case of electrodeposited (ED) multilayers with soft magnetic layers composed of Ni-Co, Fe-Co, Fe-Ni or Fe-Co-Ni alloys, a lot of efforts have already been devoted to the study of their GMR characteristics, especially in Ni-Co/Cu and Fe-Co-Ni/Cu multilayers. A critical overview of these former works has been given separately for each system in the corresponding sections of a recent review.[24] Some further reports have been published since then on the *GMR* of ED Ni-Co/Cu (Refs. 25-29), Ni-Fe/Cu (Ref. 30), Co-Fe/Cu (Ref. 31),

Co/Cu (Refs. 32-35) and Ni/Cu (Refs. 34, 36-38) multilayers.

A common feature of the majority of these works is that the multilayers investigated were prepared either by a galvanostatic/galvanostatic (G/G) or by a potentiostatic/potentiostatic (P/P) pulse combination whereby the potential of the second potentiostatic pulse applied for the deposition of the Cu layers was not optimized. It has been pointed out[39-41] that under such deposition conditions (i) the layer thicknesses will not be equal to the preset nominal values (in the G/G mode always; for too positive Cu deposition potentials in the G/P and P/P modes) but the magnetic layer thickness will be less and the Cu layer thickness will be larger or (ii) the Cu layer will be contaminated with the magnetic elements (for too negative Cu deposition potentials in the G/P and P/P modes) which is deleterious for the GMR. If the magnetic layer is strongly dissolved during the Cu deposition pulse, its thickness may reduce to the extent that some isolated regions appear in the magnetic layer which may show superparamagnetic (SPM) behavior and this leads to the appearance of a SPM contribution to the *GMR* with high saturation field.[42] Therefore, in order to properly control the individual layer thicknesses and to avoid the incorporation of magnetic elements in the Cu layer as well as to suppress the formation of SPM regions in the magnetic layers, the non-magnetic layer should be deposited at an optimized Cu deposition potential which can be established by various methods.[39,43,44]

In an effort to prepare ED multilayers with soft magnetic layers and regarding the promising sensor application potential of Ni-Co/Cu multilayers from the view-point of GMR,[18-22] we have carried out an investigation of ED $Ni_{50}Co_{50}$/Cu multilayers prepared at a Cu deposition potential optimized as described in Ref. 39. This enabled us to establish the true dependence of the *GMR* in this system on the thickness of both constituent layers as well as on the total multilayer thickness. This helps us in establishing whether an oscillatory GMR which is well-demonstrated for physically deposited Ni-Co/Cu multilayers[16-22] is exhibited also by the ED counterparts on which controversial results have been reported.[24] Since an oscillatory GMR arises due to spin-dependent scattering events between adjacent ferromagnetic layers, the suppression of the formation of SPM regions or at least the separation of their *GMR* contribution[42] is also required.

In addition, surface roughness measurements have also been performed in order to reveal the evolution of roughness with layer thicknesses and total multilayer thickness as well as the impact of roughness on GMR behavior.

## Experimental details

*ED Ni-Co/Cu multilayer preparation and characterization.* — As was in the case of depositing d.c.-plated Ni-Co alloy layers,[45] two different aqueous electrolytes were first prepared also for the electrodeposition of Ni-Co/Cu multilayers. Each electrolyte contained the sulfate of one of the two constituent magnetic metals only. The composition of the first electrolyte was 0.010 mol/ℓ $CuSO_4$, 0.10 mol/ℓ $Na_2SO_4$, 0.25 mol/ℓ $H_3BO_3$, 0.25 mol/ℓ $H_2NSO_3$ and 0.74 mol/ℓ $NiSO_4$. The last component was substituted in the second one with 0.74 mol/ℓ $CoSO_4$. The pH was set to 3.25 by adding NaOH to both solutions. The choice of this pH value was based on some preliminary experiments to get appropriate deposition conditions.[40,46] The $Ni^{2+}$- and $Co^{2+}$-containing stock solutions were mixed in various ratios to obtain electrolytes for depositing multilayers with different compositions in the magnetic layers.

The Ni-Co/Cu multilayers were deposited on a [100]-oriented, 0.26 mm thick silicon wafer covered with a 5 nm thick Cr and a 20 nm thick Cu layer both made by evaporation. The purpose of the chromium layer was to assure adhesion and the Cu-layer was used to provide the electrical conductivity of the cathode surface.

The deposition was performed in a tubular cell of 8 mm x 20 mm cross section with an upward facing cathode at the bottom of the cell.[24,40] Electrodeposition was carried out by a galvanostatic-potentiostatic (G/P) pulse combination.[24,40] For the deposition of the magnetic layer, galvanostatic (G) mode was used at $-35.0$ mA/cm$^2$ current density. At this high current density, less than 1 at.% Cu gets incorporated in the magnetic layer,[47] which does not deteriorate the magnetic and transport properties of the layer. For the Cu-layer, potentiostatic (P) mode was used at -0.585 V vs. the saturated calomel electrode (SCE) according to a previous optimization of the potential.[39]

By varying the deposition time in the G mode, the magnetic layer thickness could be set to a predetermined value. For d.c.-plated Ni-Co layers, previous profilometric measurements established[45] that the current efficiency is high enough, namely 96 %, to assume that the actual layer thicknesses are fairly close to the preset values calculated from Faraday's law.

For controlling the thickness of the Cu layer, the charge flowing through the system was measured during the P pulse. Then, from Faraday's law, one can calculate the charge necessary to get the preset layer thickness. The current efficiency for Cu deposition at the optimal potential is usually taken as 100 % since the $H_2$ evolution is negligible; therefore, we also used this value.

Due to the optimization of the Cu deposition potential, the previously deposited Ni-Co alloy layer cannot dissolve during the P pulse. It is ensured this way that both the magnetic and non-magnetic layer will have a thickness as preset from the electrodeposition parameters. It should be noted, furthermore, that our recent XRD and TEM studies indicated that under such controlled multilayer deposition conditions, the actual layer thicknesses were very close to, although slightly above the electrochemically preset values.[35,48]

Several sample series were produced with the common goal of investigating the effect of both individual layer thicknesses and total multilayer thickness on the roughness and electrical transport properties of the samples (see Table 1). Series 1 to 4 were designed to investigate the effect of the total multilayer thickness ($\Sigma d$). The thickness of the magnetic layer was fixed at 2.0 nm. For series 1 and 2, the total thickness of the multilayers were set to 50 and 100 nm, respectively, whereas the Cu layer thickness was varied between 0.8 and 9 nm. For series 3 and 4, the total thickness was set to 300 and 700 nm, respectively, and the Cu layer thicknesses were varied between 0.8 and 6 nm. For series 5, 6, 7 and 8 (where series 7 is identical with series 3, the distinction with the naming is only practical), the total multilayer thickness was fixed to 300 nm and the magnetic layer thickness was set to 1.0, 1.5, 2.0 and 2.5 nm, respectively, while the Cu layer thickness was varied again between 0.8 and 6.0 nm.

The overall multilayer composition was measured with electron probe microanalysis (EPMA) in a JEOL JSM-840 scanning electron microscope.

The root-mean-square surface roughness ($R_q$) was investigated with atomic force microscopy (AFM). The error of $R_q$ is about 0.1 nm. The Si/Cr/Cu substrate showed height fluctuations not larger than 3 nm.

*Controlling the magnetic layer composition in ED Ni-Co/Cu multilayers.* — Though Ni-Co alloys with any Ni/Co ratio can be deposited from a single electrolyte, we should keep in mind that Ni and Co show anomalous codeposition. This means that, though Ni is more noble metal than Co, the Co concentration relative to Ni in the deposit is higher than the concentration of $Co^{2+}$ ions relative to the $Ni^{2+}$ ions in the solution. As the alloy is deposited, the near-substrate region of the electrolyte becomes depleted for $Co^{2+}$ ions and thus, as the deposition proceeds, the Co-concentration of the deposit decreases[49] This leads to a concentration gradient in the magnetic layer. Therefore the concentrations of the deposited layers as a function of the concentration of the solution has to be determined.

Over the entire composition range, there is a strong correlation between the relative ion concentration of cobalt in the electrolyte and the Co-content in the deposited Ni-Co alloy layers in the multilayer obtained with a constant current density of the G pulse. However, Fig. 1 demonstrates that the steepness of the deposit composition evolution for small $Co^{2+}$ concentrations in the bath is much higher for the magnetic layers in the multilayers (full symbols) than for d.c.-plated Ni-Co alloys[45] (open symbols). This is due to the anomalous codeposition properties of Co with Ni owing to which at the beginning of the layer formation the electrolyte is depleted for $Co^{2+}$-ions. This makes the electrolyte near the sample surface to become more and more rich in $Ni^{2+}$-ions as the magnetic layer gets thicker. Thus, the regions of the deposited alloy at larger distance from the substrate become richer in Ni until a steady state is reached at a certain Ni/Co ratio in the alloy. Because pulse plating is used for preparing the multilayers, after depositing a few nanometers of the magnetic alloy layer, the G pulse ends and during the subsequent Cu deposition pulse (P) which lasts for at least 10 seconds or, sometimes, even longer than one minute, the $Co^{2+}$-ion concentration of the bath at the cathode-electrolyte interface can recover to the bulk concentration of the electrolyte existing far from the sample surface. Therefore, the next layer will grow as a Co-rich alloy again and this explains the higher Co-content of the magnetic layer in the multilayers in comparison with the d.c.-plated Ni-Co alloy deposited under identical conditions (bath concentration and deposition current density).

With the help of the composition analysis data of Fig. 1 for ED Ni-Co/Cu multilayers, we could establish the appropriate bath composition for the preparation of multilayers with approximately equal composition of Co and Ni. The solution contained 5 V/V % from the Co-electrolyte and 95 V/V % from the Ni-electrolyte which means 0.703 mol/ℓ $NiSO_4 \cdot 7 H_2O$ and 0.037 mol/ℓ $CoSO_4 \cdot 7 H_2O$ apart from the other components.

One can also see that the Co/Ni ratio in the magnetic layer does not vary noticeably with deposition current density during the G pulse. In order to keep the Cu content in the magnetic layer as low as possible, the higher current density value -35.0 mA/cm$^2$ was chosen for preparing the ED Ni-Co/Cu multilayers for the present study.

*Measurement of electrical transport properties.* — The room-temperature zero-field electrical resistivity ($\rho_0$) of the ED Ni-Co/Cu multilayers was measured in the as-deposited state of the samples, i.e., while still being on their Si/Cr/Cu substrates and before putting them in a magnetic field.

The magnetoresistance (*MR*) was measured at room temperature as a function of the external magnetic field (*H*) up to 8 kOe. The MR ratio was defined with the formula $MR(H) = [R(H) - R_0]/R_0$ where $R_0$ is the resistance maximum of the sample in a magnetic field close to zero and $R(H)$ is the resistance in an external magnetic field *H*.

The magnetoresistance was determined in the field-in-plane/current-in-plane geometry in both the longitudinal (*LMR*, magnetic field parallel to the current) and the transverse (*TMR*, field perpendicular to the current) configurations. If one takes the difference between the longitudinal and the transverse component, the anisotropic magnetoresistance can be obtained: $AMR = LMR - TMR$. For the sake of clarity of the data, only the isotropically averaged *GMR* values are plotted in the figures which quantity was determined as $GMR = (1/3) LMR + (2/3) TMR$.

The measured *MR(H)* curves were decomposed according to a procedure described previously[42] in order to establish the ferromagnetic ($GMR_{FM}$) and superparamagnetic ($GMR_{SPM}$) contributions to the *GMR*. The measured resistivity and *MR* data were always corrected for the shunting effect of the substrate.

### Results and discussion

*Surface roughness behavior.* — For some selected samples of series 1 to 4, the root-mean-square surface roughness ($R_q$) was determined (see Fig. 2). Three different Cu layer thicknesses were selected: 0.8, 3.4 and 6.0 nm whereby the magnetic layer thickness was held constant at $d_{NiCo} = 2$ nm. The data are displayed in Fig. 2 as a function of the total multilayer thickness $\Sigma d$.

The $R_q$ parameter shows an exponential increase with total multilayer thickness as usually found for layers obtained via an atom-by-atom deposition process.[50] This behavior can originate from two sources. The first is the cumulative surface roughening: the peaks at the top of the multilayer are more accessible to the ions in the solution and thus can grow faster than the valleys between them. The other is the misfit between the lattice parameters of the Ni-Co alloy and the Cu metal as a result of which a 3–D growth process tends to emerge. Once a small grain of the material of the subsequent layer can be deposited on the surface, it is energetically more favorable if the 3D-growth proceeds further instead of the formation of new "islands" of the same material (i.e., nucleation which is always hindered to some extent on the surface of a foreign metal or alloy). These grains occur at several surface sites and, after a certain time, they coalesce to form a continuous layer. By that time, at the connection points, there can only be a single atomic layer of the material deposited whereas at the point where the nucleation started, a high peak could already grow. The blocked nucleation and, thus, rather an island-like growth with coalescence is especially prone in the case of Cu layer growth on top of a Co layer as was concluded for both evaporated[51] and electrodeposited[52] Co/Cu multilayers.

The continuous multilayer roughening with increasing total multilayer thickness can be better assessed if samples are compared with different total thicknesses but with the same magnetic and non-magnetic layer thicknesses. If the topography of their surfaces is measured and the average height is set to the nominal thickness of the multilayer, the multilayer surface evolution along the thickness can be visualized as shown in Fig. 3. If we compare Figs. 3a and 3c, it can be seen that the 700-nm total thickness profile for a multilayer with 0.8-nm thick Cu layer is practically the same as that of the 300-nm total thickness profile of the multilayer having 6.0 nm Cu layer thickness. This clearly shows the roughening of the total multilayer due to the Cu layer thickness; however, this roughening effect cannot be recognized until a certain total multilayer thickness is achieved.

These surface roughness profiles show clearly that the nucleation is very uneven, high peaks and valleys develop as the total multilayer thickness increases.

For the series with 300 nm total thickness and with different magnetic and Cu layer thicknesses, $R_q$ showed a shallow maximum at about $d_{NiCo} = 1.5$ nm as a function of the magnetic layer thickness (Fig. 4). Since the Ni-Co alloy layer starts to grow in the form of separated islands, until these islands are not connected, the roughness of the layer increases. When they coalesce and start to form a continuous layer, the roughness starts to decrease until it reaches its minimal value. This effect leads to the observed variation of the $R_q$ parameter: $d_{NiCo} = 1.5$ nm is the nominal layer thickness at which the separate islands start to coalesce with each other. Growing thicker magnetic layers results in the observed decrease of the roughness.

*Zero-field electrical resistivity.* — Figure 5a shows the zero-field resistivity ($\rho_0$) data for the multilayers in series 1 to 4. The two horizontal lines show the $\rho_0$ values of the bulk $Ni_{50}Co_{50}$ alloy and bulk Cu. The thick decreasing curve shows the values given by the parallel resistance model,[53,54] by assuming that the individual layers are perfectly smooth and thus the total resistance could be calculated from their bulk resistivity and thickness as a parallel resistor system. For most of the multilayer samples, the experimental data are larger than the model values indicating the importance of interface scattering[53,54] due to the fact that the individual layer thicknesses are comparable to or even smaller than the electron mean free path in the bulk Ni-Co alloy and the Cu metal.

Since the total thickness is the same for all samples in a given series, with the increase of the Cu layer thickness, there are more Cu and less Ni-Co alloy in the multilayer, which results in the observed decrease of the resistivity of the multilayers for all total multilayer thicknesses. For large Cu layer thicknesses, the zero-field resistivity data approach, at least for the smallest total multilayer thickness, the bulk value of Cu.

Due to the nanoscale thickness of the constituent layers, most of the electron scattering occurs at the interfaces of neighboring layers and thus the more the interfaces in the multilayer, the higher the overall resistance. This effect has to be taken into account as the thickness of the Cu layer is varied while the total multilayer thickness is held constant. In this case, the resistivity decreases with decreasing $d_{Cu}$ not only because the relative amount of the smaller-resistivity multilayer component (Cu) increases in comparison with the higher-resistivity Ni-Co alloy component but also because the number of interfaces is reduced with increasing Cu layer thickness.

Furthermore, because the layers are not perfectly planar but become rougher and rougher as the total multilayer thickness increases, the probability of the occurrence of interface scattering events in the applied current-in-plane geometry also increases; this corresponds actually to the situation what is called a current-at-angle-to-plane (CAP) geometry.[55] This is the basic explanation for the increase of $\rho_0$ with total multilayer thickness. Another contributing factor can be that the upper part of a very rough multilayer does not take part in the electrical conductance, and the effective sample thickness ranges only until the sample can be taken as compact.

If the multilayer is thick enough (such as, e.g., the case is for 700 nm), the increment of the resistivity can be so high that the overall resistivity is higher than either of the bulk resistivity of the materials in the layers. Furthermore, as the Cu-layer thickness is increased within each series, the surface of the samples became rougher and rougher, which could be observed even by naked eye. This effect combined with the roughening as the total thickness increases ends up in a very sharp increase of the resistivity for series 3 and 4 ($\Sigma d = 300$ nm and 700 nm, respectively) beyond $d_{Cu} = 5$ nm (Fig. 5a). For clarity, the trend of the evolution of the data is only indicated for the series with 700 nm total thickness by the dashed line for $d_{Cu} > 5$ nm.

As be seen from Fig. 5b, the thickness of the magnetic layer has no significant effect on the resistivity. Only the high zero-field resistivity for the samples with the thinnest non-magnetic layers should be mentioned: the resistivity values lie much higher than the bulk values of the layer constituents. This is due to the microstructure of the sample: the amount of materials deposited in the subsequent pulses (both G and P) are so small that they cannot form percolating layers but remain in separate islands (see Fig. 8 in Ref. 52). The whole structure of the sample is a mixture of Ni-Co alloy and Cu metal grains. The electron scatterings at the numerous interfaces between these grains increase the total resistivity of the sample. Nevertheless, the expected continuous decrease of the resistivity towards the bulk Cu value with increasing Cu layer thickness can still be observed also here.

*Magnetoresistance results for the present ED $Ni_{50}Co_{50}$/Cu multilayers.* — For all the multilayers with $d_{Cu} = 0.8$ nm, the longitudinal MR component was positive and the transverse one was negative. This means that the dominant contribution to the observed magnetoresistance for these multilayers comes from AMR. This indicates that the magnetic layers percolate through the numerous pinholes in the not completely continuous Cu layers. The adjacent magnetic layers are connected physically and, thus, coupled ferromagnetically.[56,57] Therefore, the magnetic material does not appear in the form of separate layers but percolates as a whole and thus the sample behaves as a bulk ferromagnetic material showing only AMR.

For all other samples ($d_{Cu} > 0.8$ nm), a clear GMR behavior could be observed (both the *LMR* and *TMR* components were negative). The measured *MR(H)* curves were analyzed according to the usual procedure[42] and this way the ferromagnetic ($GMR_{FM}$) and superparamagnetic ($GMR_{SPM}$) contributions as well as the total saturation GMR ($GMR_s$) were determined. The isotropically averaged *GMR* values will only be presented which were determined from the measured *LMR* and *TMR* data as described in the experimental section.

Figure 6 shows the $GMR_{FM}$ data for series 1 to 4 as a function of the Cu layer thickness for various total multilayer thicknesses. For each series, the *GMR* increases monotonically up to 5-6 nm Cu layer thickness where it reaches a maximum and then slightly reduces. The reduction of *GMR* for large spacer layer thicknesses may partly come from a simple dilution effect. Namely, with increasing $d_{Cu}$ the bilayer repeat will be larger. In this manner, the number of FM/NM interfaces per unit thickness which are responsible for the spin-dependent scattering processes yielding the GMR effect is reduced.

It can also be inferred from Fig. 6 that the $GMR_{FM}$ component increases with total multilayer thickness from 50 nm to 300 nm and then drops for 700 nm. In order to explain the observed evolution of the GMR with total multilayer thickness, the variation of the zero-field resistivity $\rho_0$ and of the field-induced change of the electrical resistivity $\Delta\rho_H$ has to be considered separately. The *MR* definition given in the experimental section as $MR(H) = [R(H) - R_0]/R_0$ is equivalent to $MR(H) = [\rho(H) - \rho_0]/\rho_0 = \Delta\rho_H/\rho_0$. In the following, $\Delta\rho_H$ will refer to the value calculated from the $GMR_{FM}$ data by using these relations.

The evolution of $\rho_0$ with $d_{Cu}$ and total multilayer thickness was given in Fig. 5a for multilayers in series 1 to 4 whereas the $\Delta\rho_H$ data for the same samples are presented in Fig. 7. In agreement with the nearly same roughness for 50 nm and 100 nm total thickness (Fig. 2), their $\rho_0$ values are also very similar (Fig. 5a). Thus, the definitely higher $\Delta\rho_H$ values for 100 nm (Fig. 7) result in the clearly larger *GMR* values (Fig. 6). Although the roughening is more pronounced for 300 nm total thickness and there is also a substantial increase in $\rho_0$ as well, the much larger $\Delta\rho_H$ values (Fig. 7) still lead to a significant further increase of the GMR (Fig. 6). On the other hand, the drastic roughening for 700 nm total thickness leads to a very large zero-field resistivity (Fig. 5a) and even if $\Delta\rho_H$ further increases for 700 nm total thickness (Fig. 7), the large $\rho_o$ values suppresses the resulting *GMR* (Fig. 6).

Since each series can be taken as a slice from the bottom of the thicker series, if the multilayers were homogenous along their whole thickness, $\Delta\rho_H$ would be identical for all series. We should point out, on the other hand, that the continuous increase of $\Delta\rho_H$ with total multilayer thickness can be definitely related with the cumulative roughening. Namely, increasing roughening implies an enhanced undulation of the layer planes which, due to the applied current-in-plane geometry for measuring the magnetoresistance, enhances the probability of interface scattering (actually, many of the scattering events will be of the CAP type).

The $GMR_{FM}$ data for series 5 to 8 can be seen in Fig. 8. For all series, the $GMR_{FM}$ term saturates at around 4 nm Cu layer thickness whereas it hardly changes with $d_{NiCo}$ except that perhaps the *GMR* is somewhat smaller for the series with the thickest magnetic layer which can again be considered as resulting from a dilution effect. The data match fairly well the corresponding results shown in Fig. 6.

Apart from some scattered data at $d_{Cu} = 3.4$ nm, Fig. 9 shows that $\Delta\rho_H$ reduces slightly with increasing magnetic layer thickness and this is again due to a dilution effect. The shallow maximum for intermediate Cu layer thicknesses is the result of a competition between improved *GMR* with increasing spacer thickness and *GMR* reduction with increasing bilayer repeat length.

By decomposing the superparamagnetic contribution to the *GMR*, we could determine the ratio of the SPM contribution to the total measured *GMR* and the results are shown in Figs. 10 and 11. It can clearly be seen that, for thin Cu layer thicknesses, half of the observed, otherwise fairly small total *GMR* (Fig. 6) comes from consecutive electron scattering events along paths connecting an SPM and a FM region whichever is the first.[24,42] When the total *GMR* reaches a sizeable value (around 5 nm spacer layer thickness, cf. Fig. 6), then the relative importance of the SPM contribution strongly decreases and the observed *GMR* will be predominantly contributed by the FM-spacer-FM scattering events (Fig. 10).

An inspection of Fig. 11 reveals that in the investigated range of magnetic layer thicknesses (1.0 to 2.5 nm), the $GMR_{SPM}/GMR_s$ ratio does not change appreciably with magnetic layer thickness. This implies that even for the multilayers with 1 nm $Ni_{50}Co_{50}$ layer thickness, the magnetic layer is mostly ferromagnetic, i.e., mainly consists of percolating $Ni_{50}Co_{50}$ regions. Though a contribution of the SPM regions to the total measured GMR is present but apparently it is the same for all magnetic layer thicknesses. The formation mechanism of SPM regions during growth is probably the scheme suggested by Ishiji and Hashizume.[58] In this scheme, the probability of the formation of SPM regions scales with surface roughness. The present data as well our previous results[29] on another set of ED Ni-Co/Cu multilayers comply with this formation mechanism.

*Comparison of GMR data with previous results for Ni-Co/Cu multilayers.* — As noticed in the Introduction, physically deposited FM/NM multilayers usually exhibit an oscillatory behavior of the GMR as a function of the non-magnetic spacer layer thickness. This feature has been well documented for sputtered Ni-Co/Cu multilayers over the whole composition range of the magnetic layer from pure Ni to pure Co (Refs. 18, 21 and 22). Pronounced GMR maxima appear at about $d_{Cu} = 0.9$ nm, 2.0 nm and 3.5 nm. The positions of these maxima correspond to spacer layer thicknesses where an AF coupling ensures that the adjacent magnetic layers have an antiparallel orientation with respect to each other in zero magnetic field and this state is associated with a high electrical resistivity. As a consequence of this zero-field antiparallel alignment, the resistivity change upon the application of a sufficiently high magnetic field ($H_s$) forcing all the layer magnetizations to take a parallel alignment will be large, yielding the *GMR* maximum. At intermediate spacer thicknesses (e.g., 1.5 nm and 2.5 nm), the spacer-mediated exchange coupling between adjacent magnetic layers is ferromagnetic, i.e., all the layer magnetizations are parallel and, therefore, they are in a low-resistance state already in zero magnetic field and such multilayers do not exhibit a GMR effect. These spacer thicknesses correspond to *GMR* minima.

In line with our previous conclusions that both ED Co/Cu (Ref. 59) and Ni/Cu (Ref. 38) multilayers do not exhibit an oscillatory GMR behavior, the present study demonstrated (Figs. 6 and 8) that the situation is the same also for ED $Ni_{50}Co_{50}$/Cu multilayers. It should be emphasized that when looking for an oscillatory GMR behavior, only the $GMR_{FM}$ component should be considered as a function of the spacer layer thickness. This is because the oscillatory behavior is due to an oscillation of the sign of the interlayer exchange coupling and such a coupling can only occur if the adjacent magnetic layers are in FM state since such a coupling can not realize between FM and SPM regions. It is also an important aspect that the actual layer thicknesses should be fairly close to the preset values which, as indicated in the Introduction, is not always properly fulfilled under certain electrochemical conditions during multilayer preparation. These are the main guidelines along which the reported studies on the spacer layer thickness dependence of *GMR* in ED Ni-Co/Cu multilayers will be evaluated in the following.

Although there have been numerous reports on the spacer layer thickness dependence of ED Ni-Co/Cu multilayers,[60-70] the results are rather controversial. This is partly connected with the fact that in many cases only the *GMR* value measured in the highest available magnetic field was displayed against $d_{Cu}$ whereas the *MR(H)* curves if presented at all often indicated a significant $GMR_{SPM}$ component in the measured magnetoresistance, at least for certain ranges of $d_{Cu}$. Another problem is that the deposition conditions specified often indicated that a significant magnetic layer dissolution may have occurred during the Cu deposition cycle as a result of which the actual Cu layer thicknesses are larger by an undefined amount than the preset nominal values due to the usually applied coulometric control of layer thicknesses. The magnetic layer composition (Co:Ni ratio and/or amount of Cu incorporated in the magnetic layer) also varied from study to study if it was specified at all.

Irrespective of all these uncertainties, qualitatively very similar results to our data (see Figs. 6 and 8) were reported for ED Ni-Co/Cu multilayers in Refs. 64 and 66-69 in that the *GMR* magnitude increased monotonically with $d_{Cu}$ up to a certain Cu layer thickness and then it either remained constant or slightly decreased for thicker Cu layers. The position of the *GMR* maximum or the $d_{Cu}$ value at the onset of *GMR* saturation varied in the range 2 nm to 7 nm. It should be noted that in each of the above reports, the presented *MR*(*H*) curves indicated that for multilayers with the largest *GMR* values (i.e., for fairly large $d_{Cu}$ values), the $GMR_{FM}$ term was the dominant one (manifested by the low saturation fields around 1 to 2 kOe). On the other hand, for low Cu layer thicknesses (typically around 1 nm), the measured *MR*(*H*) could be assessed as being dominated by the $GMR_{SPM}$ contribution (as found also in the present work, see Figs. 7 and 9). The results of these previous reports[64,66-69] firmly support our finding about the lack of an oscillatory GMR behavior in ED Ni-Co/Cu multilayers. The highest possible *GMR* values achieved in the individual studies were rather scattered and varied according to the specific deposition conditions (deposition mode, substrate material and orientation, pH, stagnant or flow electrolyte, etc.).

Of the other reports not in conformity with our finding, we mention first the work of Alper et al.[60] in which a monotonic decrease of the *GMR* values measured at $H = 8$ kOe was obtained as $d_{Cu}$ varied from 0.7 nm to about 2.5 nm and then the *GMR* remained approximately constant or a slight increase occurred up to 3.5 nm. The *MR*(*H*) curve presented for the lowest spacer layer thickness unambiguously demonstrates that the GMR is dominated here by a $GMR_{SPM}$ contribution. It has to be also noted that the *GMR* values reported are not saturation values; hence, the change in the size of the SPM regions may also influence the shape of the *MR*(*H*) curves, which makes the comparison of single-field *GMR* values even more difficult. This dominance of the $GMR_{SPM}$ term is in line with our finding at low Cu layer thicknesses, however, the magnitude of the *GMR* is much larger in Ref. 60. We may speculate that at the applied Cu deposition potential of Alper et al.,[60] a strong dissolution of the magnetic layer may have led to a fragmentation of the magnetic layer which finally consisted of mainly SPM regions. This would correspond roughly to the situation sketched in Fig. 8 of the paper by Liu et al.[52] where it was found that when the thickness of both the magnetic and non-magnetic layers is below a certain threshold, the deposit appears as a granular material with many SPM and some FM regions embedded in the Cu matrix. Due to the granular nature of the deposit, there are numerous electron pathways between neighboring magnetic regions through the spacer and this enhances the observed *GMR* with respect to a regular, well-defined FM/NM layered structure. We believe that the same explanation applies also for the occurrence of an apparently large *GMR* maximum at $d_{Cu} = 0.7$ nm in ED Ni-Co/Cu multilayers deposited on Cu(100) and Cu(111) single-crystal substrates.[63] These latter authors have found, nevertheless, a small second *GMR* maximum at about $d_{Cu} = 1.8$ nm for multilayers grown on Cu(100) single crystals whereas no second maximum was found for multilayers grown on Cu(111) single crystals. This difference between the two multilayer sets was explained by the authors with the differences in the microstructure and this indeed seems to be a reasonable explanation. Nevertheless, we believe that the second *GMR* maximum observed for multilayers on Cu(100) single crystals may well correspond to the *GMR* maximum in the previously discussed reports (present work and Refs. 64 and 66-69) if we assume that due to the applied too positive Cu deposition potentials in Ref. 63, the actual Cu layer thicknesses are larger than the specified nominal values.

A very peculiar case is the result reported by Hua and coworkers.[61,62] They found a small *GMR* maximum at $d_{Cu}$ = 1 nm and a much larger one at $d_{Cu}$ = 2.3 nm with definitely smaller *GMR* values in between. It can be inferred from the reported *MR*(*H*) curves[61,62] that at $d_{Cu}$ = 1 nm the *GMR* is predominantly contributed to by the $GMR_{SPM}$ term and by the $GMR_{FM}$ term for $d_{Cu}$ = 2.3 nm. Thus, by considering also the very positive Cu deposition potential applied by Hua and coworkers,[61,62] the situation is very similar to the finding in Ref. 63, apart from the magnitude of the *GMR* values at around $d_{Cu}$ = 1 nm. Apparently, the fine details of the deposition conditions used by Hua and coworkers[61,62] led to a different microstructure at low Cu layer thicknesses whereas for larger Cu layer thicknesses, a behavior as observed also in the present paper was obtained.

As to the work by Kainuma et al.,[65] important details of the preparation conditions are missing from their paper; nevertheless, they reported 5 to 10 % *GMR* values for Cu layer thicknesses between 5 and 10 nm. However, in lack of presented *MR*(*H*) curves, we cannot assess the importance of the $GMR_{SPM}$ term. The reported *GMR* oscillations cannot be considered as real (especially because their positions do not correspond to the expected ones) but the overall evolution of the *GMR* is qualitatively similar to our data in Figs. 6 and 8.

Finally, we mention the work of Dulal and Charles[70] who reported *GMR* values between 0.5 and 1 % for $d_{Cu}$ ranging from 2 to 10 nm. They applied a citrate type bath at pH 6 and these conditions can explain the typical non-saturating *MR*(*H*) curves with dominant $GMR_{SPM}$ term and small *GMR* values as reported also for ED Ni-Cu/Cu multilayers.[71] Therefore, the evolution of the *GMR* with $d_{Cu}$ cannot be considered as relevant for the presence of any GMR oscillation.

**Summary**

In the present work, the surface roughness and the room-temperature electrical transport properties (zero-field resistivity and magnetoresistance) were investigated for ED $Ni_{50}Co_{50}$/Cu multilayers. For this purpose, multilayer series with 50, 100, 300 and 700 nm total thicknesses and with 1.0, 1.5, 2.0, and 2.5 nm magnetic layer thicknesses were prepared for various Cu layer thicknesses from 0.8 nm to 9 nm by electrodeposition on Si wafers with evaporated Cr and Cu underlayers.

A roughening effect with the increase of both the total multilayer thickness and the Cu layer thickness was found. A continuous magnetic layer formation could be assessed for Co layer thicknesses as high as at least 1.5 nm.

The room-temperature zero-field resistivity was found (i) to decrease with Cu layer thickness for fixed magnetic layer thicknesses and (ii) to increase as the total multilayer thickness increased. If the total multilayer thickness was held constant and the magnetic layer thickness was varied, a continuous decrease was found with increasing the Cu layer thickness whereas the magnetic layer thickness showed no influence on the resistivity.

The *GMR* was found to show (i) a maximum at $d_{Cu} = 5$ nm if the magnetic layer thickness was held constant and (ii) an increase until a certain total multilayer thickness was reached.

By properly decomposing the *GMR* into FM and SPM contributions, it could be concluded that the $GMR_{FM}$ contribution does not exhibit an oscillatory GMR in ED Ni-Co/Cu multilayers.

**Acknowledgements**

This work was supported by the Hungarian Scientific Research Fund through grant OTKA K 75008. The authors also acknowledge G. Molnár (Institute for Technical Physics and Materials Science, Research Centre for Natural Sciences, HAS) for preparing the evaporated underlayers on the Si substrates.

| Ni-Co layer thickness: 2.0 nm | Cu layer thickness (nm) | | | | | | | | | |
|---|---|---|---|---|---|---|---|---|---|---|
| | 0.8 | 1.6 | 2.4 | 3.4 | 4.0 | 5.0 | 6.0 | 7.0 | 8.0 | 9.0 |
| total thickness 50 nm | Series 1 | | | | | | | | | |
| total thickness 100 nm | Series 2 | | | | | | | | | |
| total thickness 300 nm | Series 3 | | | | | | | | | |
| total thickness 700 nm | Series 4 | | | | | | | | | |

| total thickness: 300 nm | Cu layer thickness (nm) | | | | | | |
|---|---|---|---|---|---|---|---|
| | 0.8 | 1.6 | 2.4 | 3.4 | 4.0 | 5.0 | 6.0 |
| Ni-Co layer thickness 1.0 nm | Series 5 | | | | | | |
| Ni-Co layer thickness 1.5 nm | Series 6 | | | | | | |
| Ni-Co layer thickness 2.0 nm | Series 7 | | | | | | |
| Ni-Co layer thickness 2.5 nm | Series 8 | | | | | | |

*Table 1.* Sample properties (magnetic and non-magnetic layer thickness, total thickness) for all numbered sample series. Series 3 and 7 are identical but they are referred to here with different numbers in the two sets of samples for the sake of convenience.

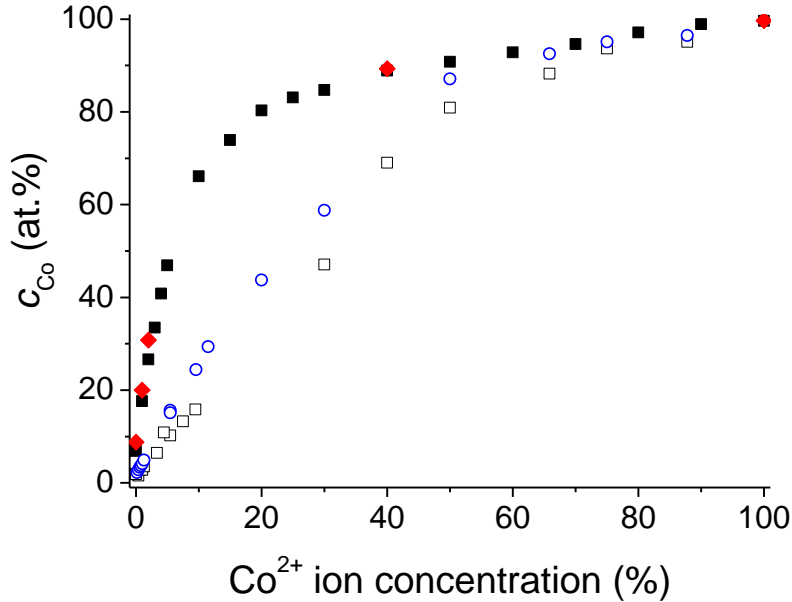

*Fig. 1* Co-content $c_{Co} = 100 \times x_{Co}/(x_{Co} + x_{Ni})$, where $x_{Co}$ and $x_{Ni}$ are the molar fractions of Co and Ni, respectively, in the ED Ni-Co/Cu multilayer deposits for current densities -35 mA/cm$^2$ (■) and -17.5 mA/cm$^2$ (◆) as a function of the relative Co$^{2+}$ ion concentration in the solution, the latter quantity defined as $100 \times c(Co^{2+})/[c(Co^{2+}) + c(Ni^{2+})]$. The open symbols (□ and ○) are the composition data for d.c.-plated Ni-Co bulk alloys[45] prepared at current densities (-31.3 mA/cm$^2$ and -18.8 mA/cm$^2$) very close to that used for preparing the magnetic layers in the multilayers.

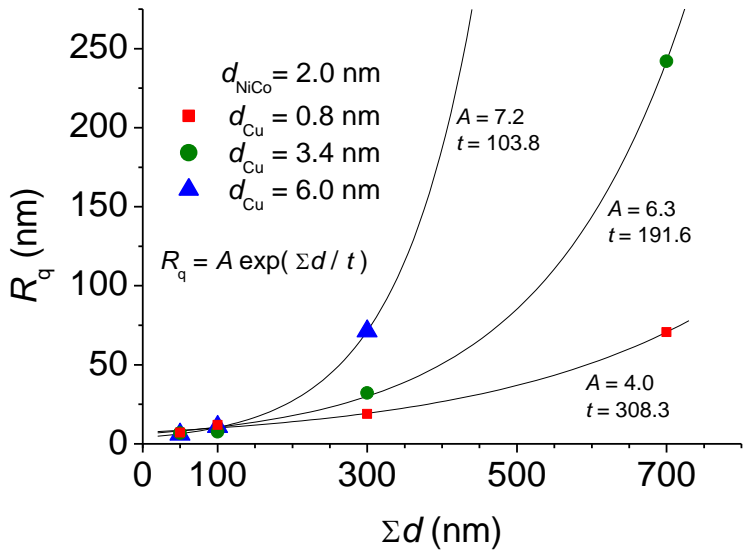

*Fig. 2* Root-mean-square roughness $R_q$ of selected Ni-Co/Cu multilayers as a function of the total multilayer thickness $\Sigma d$ for various Cu layer thicknesses and with the thickness of the FM layer fixed at 2.0 nm. The values of $A$ and $t$ correspond to the parameters resulting from an exponential fit according to the function $R_q = A \exp(\Sigma d/t)$.

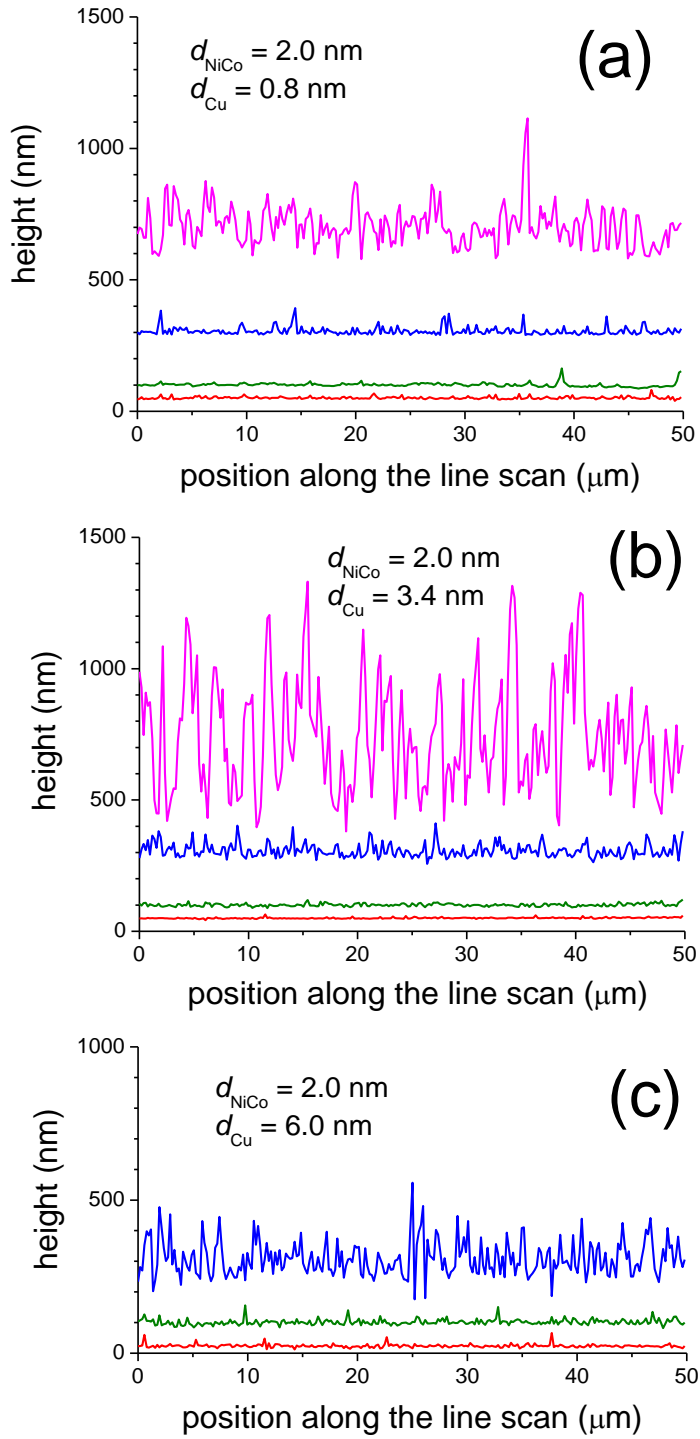

*Fig. 3* Surface profiles (line scans) of selected Ni-Co/Cu multilayers. Each panel contains line scan profiles of samples with the same magnetic and non-magnetic thicknesses but with different total thicknesses. From the bottom to the top in each of the panels, the total nominal multilayer thicknesses ($\Sigma d$) are 50 nm (red), 100 nm (green), 300 nm (blue), 700 nm (purple). The average of the height fluctuation profiles was set to the total value of $\Sigma d$. The surface of the 700 nm thick multilayer in (c) appeared so rough that it could not be measured with AFM. For $\Sigma d = 50$ and 100 nm, the $R_q$ values are practically the same for all three Cu layer thicknesses. For $\Sigma d = 300$ and 700 nm, the surface roughness is much larger and increases strongly with both Cu layer thickness and total multilayer thickness (cf. Fig. 3a).

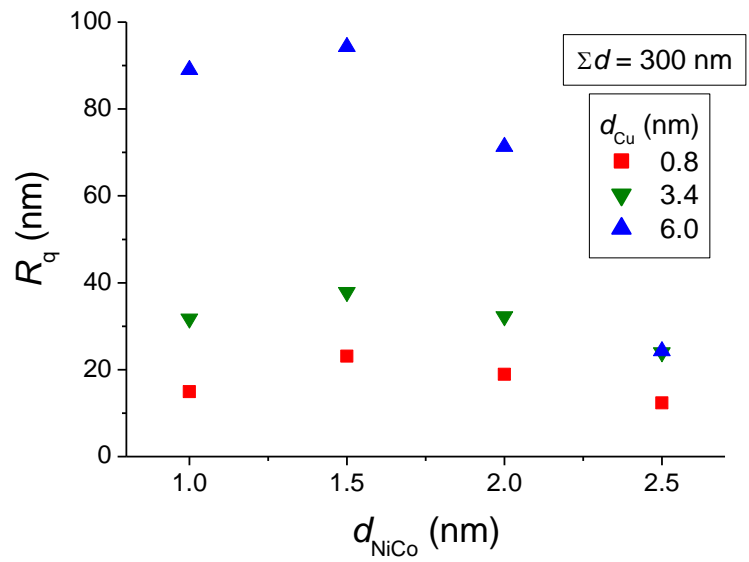

*Fig. 4* Root-mean-square roughness $R_q$ of selected Ni-Co/Cu multilayers as a function of the FM layer thickness $d_{NiCo}$ for various Cu layer thicknesses with the total multilayer thickness fixed at 300 nm.

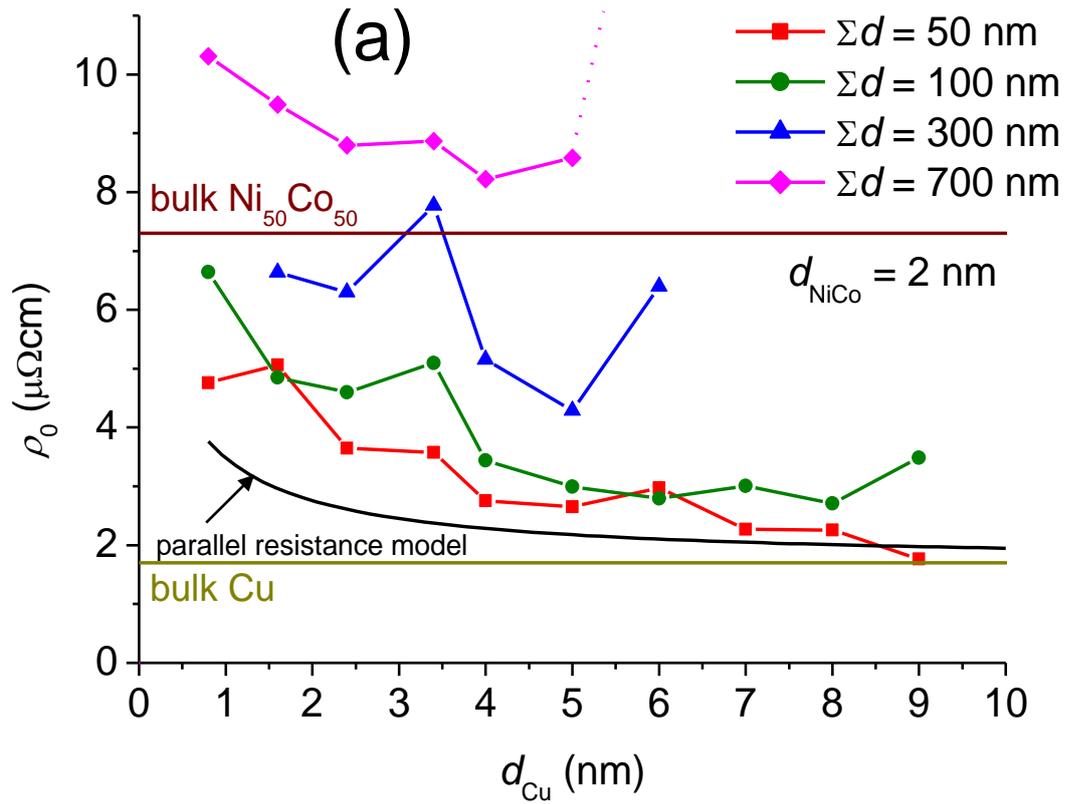

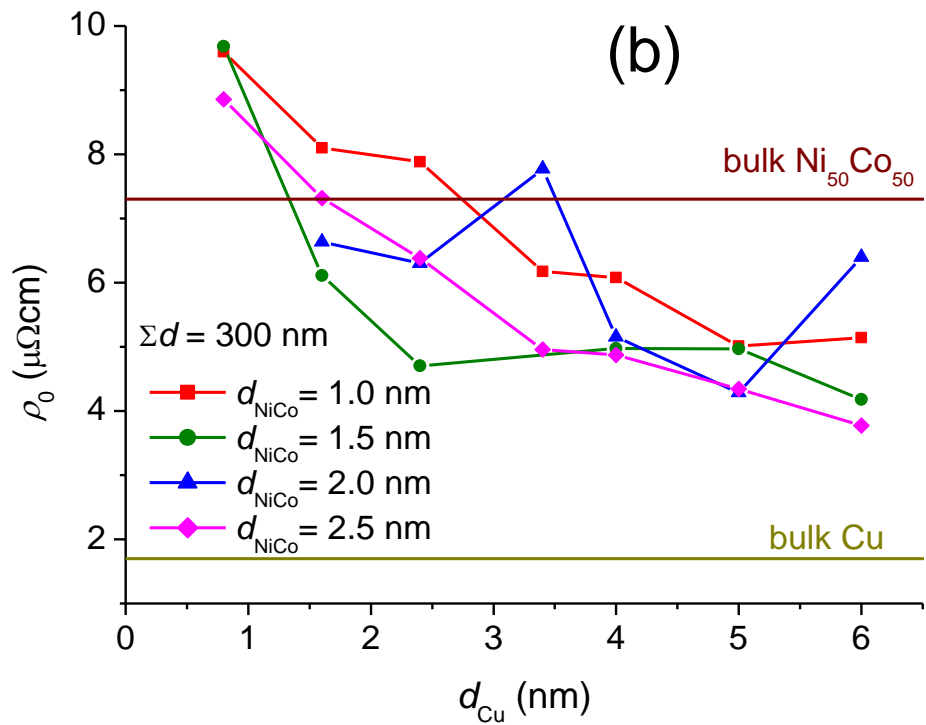

*Fig. 5* Evolution of the room-temperature zero-field resistivity with Cu layer thickness for series (a) 1 to 4 and (b) 5 to 8.

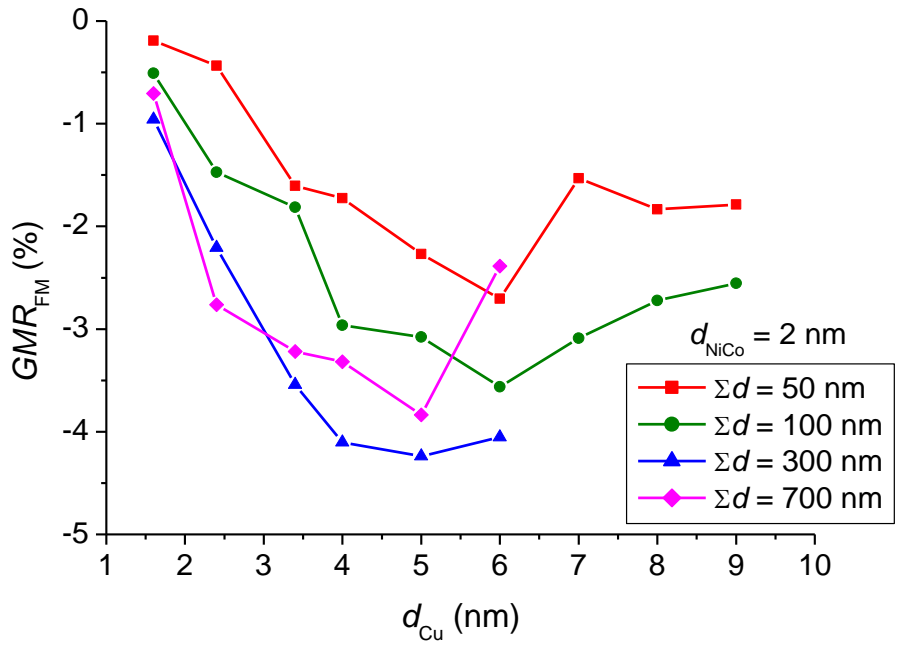

*Fig. 6* Evolution of the *GMR*$_{FM}$ contribution with Cu layer thickness for series 1 to 4 with various total multilayer thicknesses as indicated in the legend. The magnetic layer thickness was fixed at 2.0 nm.

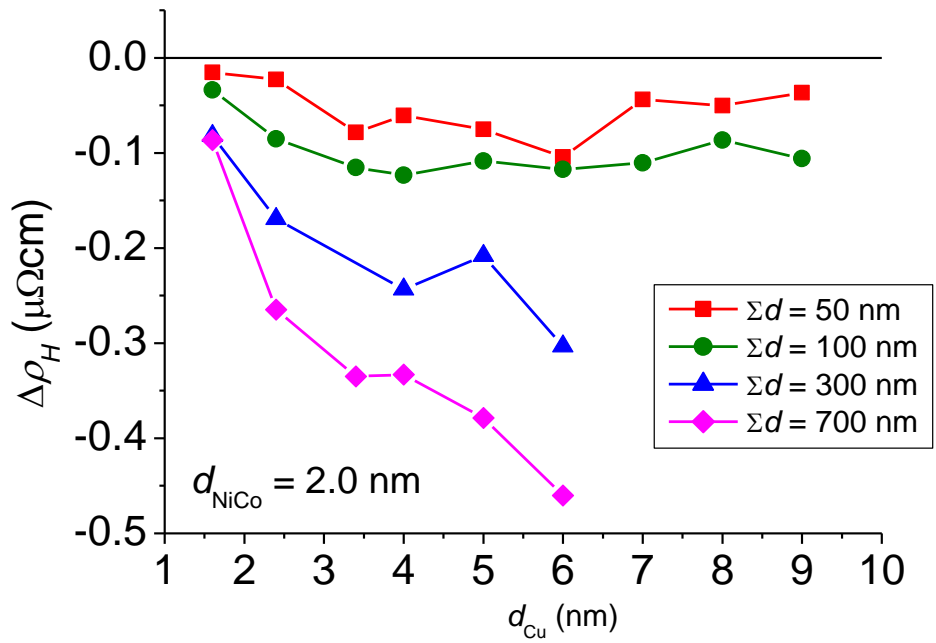

*Fig. 7* Evolution of the isotropic saturation resistivity change due to magnetic field with Cu layer thickness for series 1 to 4 with various total multilayer thicknesses as indicated in the legend. The magnetic layer thickness was fixed at 2.0 nm.

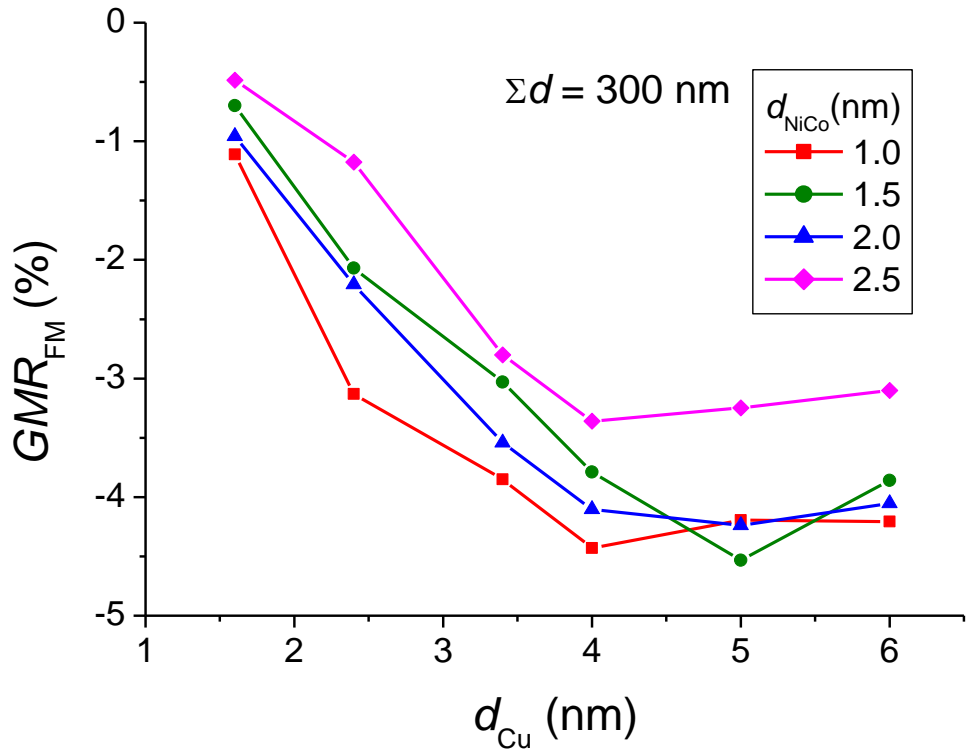

*Fig. 8* Evolution of the *GMR*<sub>FM</sub> contribution with Cu layer thickness for series 5 to 8 with various magnetic layer thicknesses as indicated in the legend. The total multilayer thickness was fixed at 300 nm.

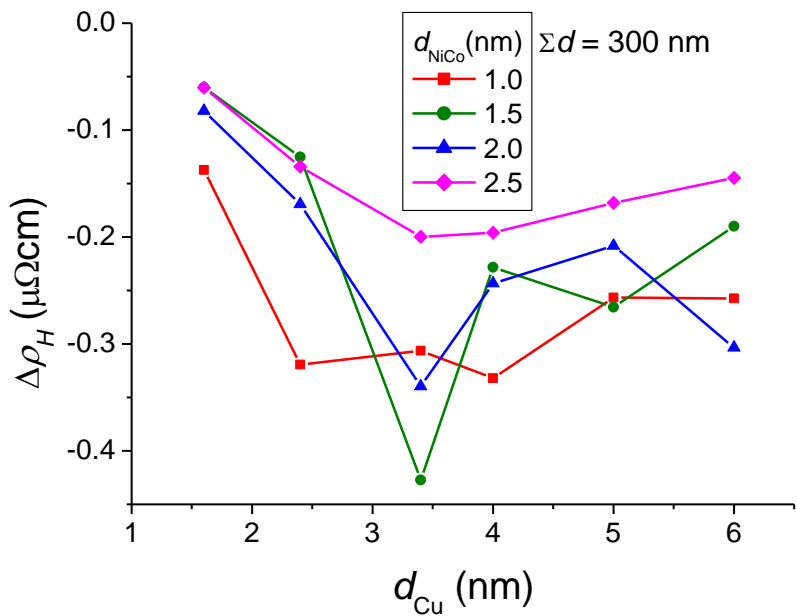

*Fig. 9* Evolution of the isotropic saturation resistivity change due to magnetic field with Cu layer thickness for series 5 to 8 with various magnetic layer thicknesses as indicated in the legend. The total multilayer thickness was fixed at 300 nm.

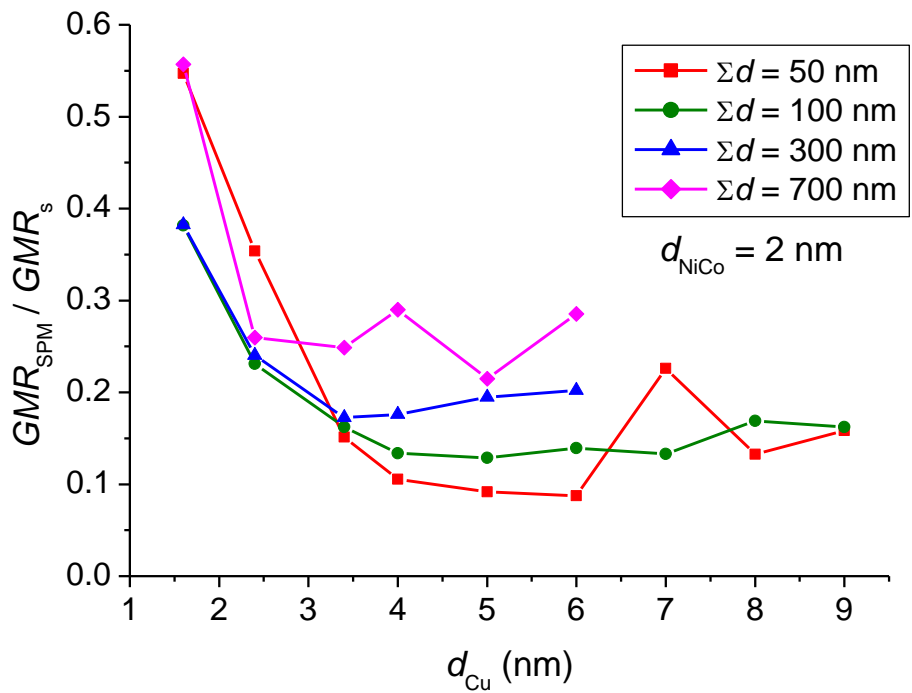

*Fig. 10* Evolution of the ratio of the GMR$_{SPM}$ contribution to the total saturation GMR (*GMR*$_s$) with Cu layer thickness for series 1 to 4 with various total multilayer thicknesses as indicated in the legend. The magnetic layer thickness was fixed at 2.0 nm.

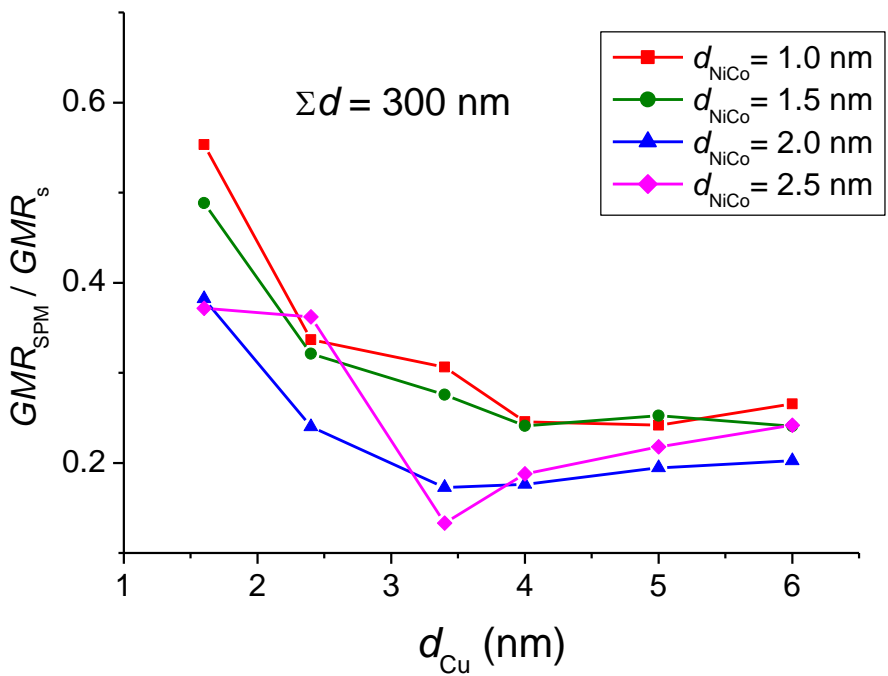

*Fig. 11* Evolution of the ratio of the *GMR*$_{SPM}$ contribution to the total GMR with Cu layer thickness for series 5 to 8 with various magnetic layer thicknesses as indicated in the legend. The total multilayer thickness was fixed at 300 nm.